\documentclass[aip,rsi,preprint,linenumbers]{revtex4-1}
\usepackage{graphicx}

\begin{document}

\title{New experimental setup for in situ measurement of slow ion induced sputtering} 

\author{P. Salou}
\author{H. Lebius}
\author{A. Benyagoub}
\author{T. Langlinay}
\author{D. Leli\`evre}
\author{B. Ban-d'Etat}
\email[]{bandetat@ganil.fr}

\affiliation{CIMAP (CEA-CNRS-ENSICAEN-UCBN), Boulevard Henri Becquerel, BP 5133, 14070 Caen Cedex 5, France}

\begin{abstract}
A new experimental equipment allowing to study the sputtering induced by ion beam irradiation is presented. The sputtered particles are collected on a catcher which is analyzed \textit{in situ} by Auger electron spectroscopy without breaking the ultra high vacuum (less than 10$^{-9}$mbar), avoiding thus any problem linked to possible contamination. This method allows to measure the angular distribution of sputtering yield. Thanks to this new setup it is now possible to study the sputtering of many elements especially light elements such as carbon based materials. Preliminary results are presented in the case of highly oriented pyrolytic graphite and tungsten irradiated by an Ar$^{+}$ beam at respectively 2.8 keV and 7 keV.
\end{abstract}

\pacs{79.20.Rf; 34.35.+a; 82.80.Pv}

\maketitle

\section{Introduction}

Sputtering has been studied for decades, this phenomenon is present in different fields of science and technology such as the cleaning of surfaces, the secondary ion mass spectrometry or the deposition of thin films\cite{Smentkowski20001}.

Sputtering is also an important parameter for the fusion technology. The walls of tokamaks are experiencing an influx of heat and particles from the plasma. The ensuing erosion, which is partially enhanced by chemical reactions, leads not only to a surface modification of the wall elements, but also to the introduction of high-Z materials into the plasma which can contribute to its instabilities. Understanding the reaction between the plasma and the wall materials is thus one of the challenges of the present fusion research\cite{Roth20091}.

Most of actual tokamak reactors are using carbon based components as wall materials\cite{Pitts2005}. This choice is mainly due to their high thermal conductivity, their low Z-value and their capability to sublimate instead of melting. However, the hydrogen isotope retention of carbon materials is high compared to other vessel materials like W and Be. In order to respect the tritium inventory limitation\cite{Federici2001}, the use of carbon based components in ITER will be restricted only to the deuterium-deuterium fusion reaction. Nevertheless, none of the above-mentioned materials reunites all the advantages of carbon. Therefore, the search for replacement materials is still the subject of intense investigations. The first iteration of wall cladding of the ITER reactor still contains carbon for divertor materials, and a total carbon-free cladding is not planned before 2020. Therefore, besides the search of alternatives, the knowledge on carbon based materials and their interaction with the plasma, in particular sputtering, needs to be improved.

Many previous studies on carbon sputtering were performed using plasma immersion\cite{Goebel198761}. Their application is however limited to basic plasmas, which do not simulate very well the plasma in tokamaks, especially in terms of energy density. An efficient way to study plasma-wall interaction is to simulate the plasma influx by means of ion beams. Contrary to the case of plasma immersion, particle properties are well defined, so the effects of ion species, charge state, energy and flux can be studied separately. It became then possible to extrapolate specific plasma properties. By using ion beam, the erosion can be characterized by the mass loss measured by quartz crystal microbalance\cite{Aumayr2004}. The sputtered particles can also be analyzed by different techniques such as mass spectrometry\cite{Keim2011}, or the so-called catcher method\cite{Bouffard1998372,Etat2004,Tripathi2003,Tripathi2008}. This last technique has the capability to study neutral species which represent the majority of the sputtered particles. 

The catcher method was adopted in this work in order to study sputtering by ion beams, specially the sputtering of fusion relevant materials such as carbon based materials. To avoid pollution, like from hydrocarbons present in air, our setup was designed in order to allow an \textit{in situ} analysis of the catcher by Auger electron spectroscopy (AES), in such way that the catcher does not leave the vacuum in-between preparation, irradiation and measurement.

\section{Experimental Setup}
\begin{figure}
	    \includegraphics[scale=1]{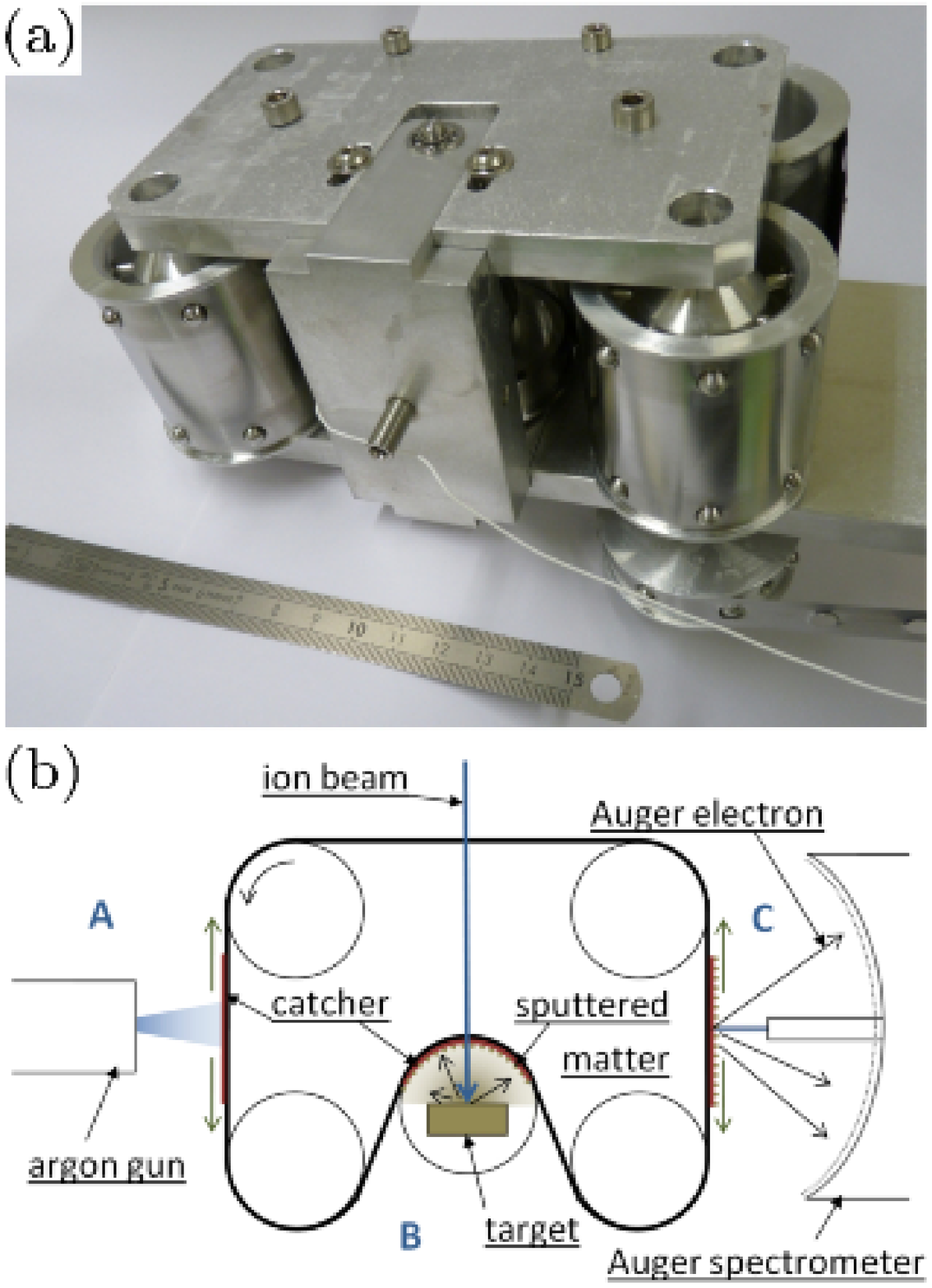}
 		\caption{%
	        Overview of the experimental setup : (a) shows the back side of the experimental setup, the beam arrives from the top of the picture, (b) shows a schematic view from the top of the setup.
	     }%
	   \label{figure1}
 \end{figure}

To measure the sputtering yield of graphite induced by ion-surface interactions using the catcher method, the catcher has to stay under ultra high vacuum during all the phases of the experiment. This means that the catcher cleaning by sputtering (Step 1), the irradiation of the target (Step 2) and the catcher analysis (Step 3) have to be performed in the same chamber without breaking the vacuum, implying that the catcher needs to be transferred to different locations. To ensure such a transfer, the catcher is stuck with UHV patches on a belt which can be moved by pulleys as shown in figure \ref{figure1}. Two of them have teeth, allowing to stretch the belt in the vertical way and to rotate it. The catcher can thereby move, facing the different areas (as shown in figure \ref{figure1}). In the area A, the catcher is placed in front of an ion sputter gun for the purpose of cleaning. In the area B dedicated to the irradiation itself, the catcher is bent around the target in order to collect all the sputtered particles produced by ion-surface interaction. Finally the catcher is analyzed in the area C by Auger electron spectroscopy. These different steps are described below.

The set-up is placed in a high vacuum chamber, pumped by a dry primary pump and two turbo-molecular pumps in series. After baking at 150$^\circ$C a pressure below 5.$10^{-10}$mbar is reached. A gate valve and a collimator are placed between the chamber and the beam line.

To collect the sputtered graphite the chosen material for the catcher is silicon, because of its good sticking coefficient, close to one\cite{Roth2004}.The silicon also allows the analysis by Auger Electron Spectroscopy (AES) since the Auger electron energies of silicon, carbon and oxygen (belonging to the native oxide) are well separated. Moreover, it is possible to get relatively flat Si surfaces and the diffusion coefficient of carbon at this surface is well known\cite{Pellerin1979}. The catcher can be made of a thin wafer of silicon, less than 40 $\mu$m thick, which can easily be bent. It is then composed of two pieces of 25 $\times$ 5 mm$^{2}$ each. The catcher can also be made of several pieces of thicker wafer.

The catcher analysis has to be sensitive to very low matter deposition, it is therefore perfectly clean before the sputtering experiment and the analysis by AES which is sensitive to sub-layer deposition. To get an atomically clean surface, a first chemical cleaning is done, the Si catcher is dived for thirty minutes into solvents bathes, respectively trichloroethylene, acetone, methanol and then in pure water. The first bath dissolves grease and hydrocarbons while the other bathes dissolve the previous solvent. A last cleaning is done under vacuum by sputtering using an argon sputter gun. The latter produces single charged ions at 500 eV, it is placed at 10 cm from the catcher and at 22 $^\circ$ from its normal. The ion beam spot on the catcher is estimated to have a FWHM of 1 cm for an intensity of 0.1 $\mu$A. This last cleaning step is carried on until no pollution is detected by the Auger Spectrometer.

\begin{figure}
 \includegraphics[scale=1]{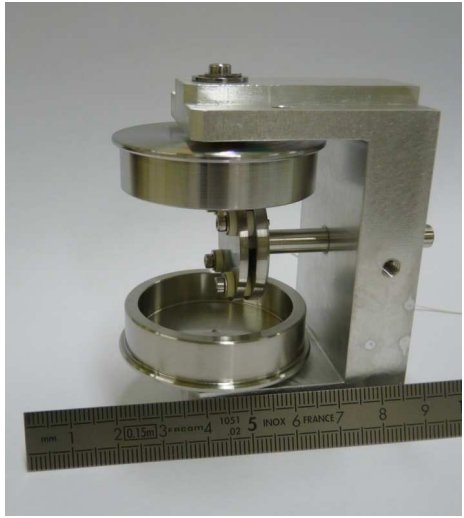}
 \caption{Target holder.\label{fig3}}
 \end{figure}

Once cleaned, the catcher is transferred to the irradiation area where it is bent around the target in a semi-circular geometry. To get the right belt curvature, an empty pulley, as shown in figure \ref{fig3}, is located in this area. The target is placed in the middle of this pulley and the catcher is positioned at 22 mm from the ion beam impact. The target holder is isolated, so by applying a positive potential to it during ion irradiation, the secondary electrons emitted from the target by incident ions are attracted back to the target allowing thus a precise measurement of the ion beam current. The target holder can host samples with a size up to 25 $\times$ 25 $\times$ 20 mm$^{3}$, allowing the study of actual plasma facing materials such as tokamak CFC tiles. The ion beam hits the target after going across the belt through 8 mm and 3 mm diameter holes. Just behind the 8 mm hole, a 5 mm and a 4 mm collimators are placed in order to measure the beam current. The target holder can be replaced by a small Faraday cup equipped with a guard ring. For the study of plasma relevant ion beams, noble gases such as He and Ar can be produced by means of filament ion sources. For other elements as H, C, N and O, electron cyclotron resonance sources, such as the Mono1000\cite{Pierret2008}, can be used. To simulate the very low energy particles present in the plasma, this setup can easily be moved to dedicated facilities equipped with low energy beam lines\cite{Lebius2003} or deceleration lenses\cite{Keim2011}.

During the irradiation, the sputtered particles are collected on the catcher. Once the desired fluence is reached, the catcher is transferred in front of the Auger spectrometer for analysis. The Auger spectrometer is a four-grid retarding field analyzer coupled with a lock-in amplifier (LEED/Auger, OMICRON). The electron beam spot diameter is approximately 0.5 mm. The catcher is scanned all along its length in order to obtain the angular sputtering yield. To deduce the amount of sputtered carbon from the Auger spectrum, we use a prior calibration of the Auger spectrometer which is described in the following paragraph. 

\section{AES Calibration}
Many theoretical models predict the evolution of Auger spectrum as a function of the material composition. In the case of a catcher material collecting another different material we can simulate it by a heterogeneous sample, composed of a substrate of silicon with an upper layer of deposited material. The transition between the condensate and the substrate is assumed sharp. As the analyzing electron spot is rather small, the material can be considered homogeneous all along the surface. Briggs and Seah\cite{Briggs1983} proposed a simple model to predict the Auger electron intensity of the different elements of the material. In this model the deposited layer is considered flat and uniform. It is thus not completely suited to describe sub-monolayer and film growth modes like that of Volmer-Weber or Stranski-Kratanov\cite{Argile1989}. However it is well adapted to the description of amorphous carbon film growth. For a condensate $C$ deposited on a substrate $S$, the amplitude of the Auger electron peak $I_{C}$ and that of the elastic electron peak $I_{EP}$ follow respectively Eqs. \ref{equation1} and \ref{equation2}:

\begin{equation}\label{equation1} 
I_{C} = k_{C}I_{0}\frac{1+r_{C}}{1+r_{S}}\lbrace1-\exp\left[-d/\Lambda_{C,E_{A}}\right]\rbrace,
\end{equation}

\begin{equation}\label{equation2} 
I_{EP} = k_{EP}I_{0}\lbrace R_{C}-(R_{C}-R_{S})\exp\left[-2d/\Lambda_{C,E_{EP}}\right]\rbrace.
\end{equation}

In these equations $I_{0}$ is the primary electron beam intensity and $d$ is the layer thickness. The constants $k_{C}$ and $k_{EP}$ depend, among other parameters, on the detector efficiency, the conversion between the number of electrons and the peak amplitude, and in the case of Auger electrons on the ionization cross section and the emission probability. The constants $\Lambda_{C,E_{A}}$ and $\Lambda_{C,E_{EP}}$ correspond respectively to the attenuation length of the electrons in the condensate at the energy of $E_{A}$ and $E_{EP}$; these parameters are linked to the inelastic mean free path (imfp) ($\lambda_{C,E_{E_A}}$ and $\lambda_{C,E_{EP}}$) through a correction factor, $k$, taking into account the elastic collisions and the detector geometry. $r_{S}$ and $r_{C}$ correspond respectively to the backscattering correction factor of the $E_{A}$ energy electrons in the material substrate and condensate.  $R_{C}$ and $R_{S}$ correspond to the backscattering probability of primary electrons on the condensate and
the substrate.

In the case of a thin film of carbon covering silicon, $r_{C}$ and $r_{S}$ are negligible. To eliminate any fluctuation of $I_{0}$, it is possible to study the ratio between the Auger and the elastic peaks. This ratio is given by the following expression:
\begin{equation}\label{equation3} 
\frac{I_{C}}{I_{EP}}= \alpha\frac{1-\exp(-d/\Lambda_{C,E_{A}})}{R_{C} - (R_{C}-R_{S})\exp(-2d/\Lambda_{C,E_{EP}})},
\end{equation}
where $\alpha=\left[k_{C}\left(1+r_{C}\right)\right]/ \left[k_{EP}\left(1+r_{S}\right)\right]$.

The above equations include intrinsic and extrinsic parameters. While intrinsic parameters can easily be found in the literature, the extrinsic parameters which are linked to the Auger spectrometer need to be determined. To measure them we used a setup (described by Akc\"{o}eltekin \textit{et al.}\cite{Akcoeltekin2009}) dedicated to sample preparation and post-irradiation analysis. It is equipped for producing clean surface samples, depositing thin calibrated carbon films, and analyzing them with the Auger spectrometer.

The thin films of carbon were deposited from a graphite rod sublimated by electron heating. The flux of evaporated particles was measured by using a quartz crystal microbalance located at the sample position. The microbalance was temperature stabilized by water cooling. During the deposition the stability of the carbon flux was monitored with the measurement of the intensity of the particle beam.

\begin{figure}
	\centering
               \includegraphics[scale=1]{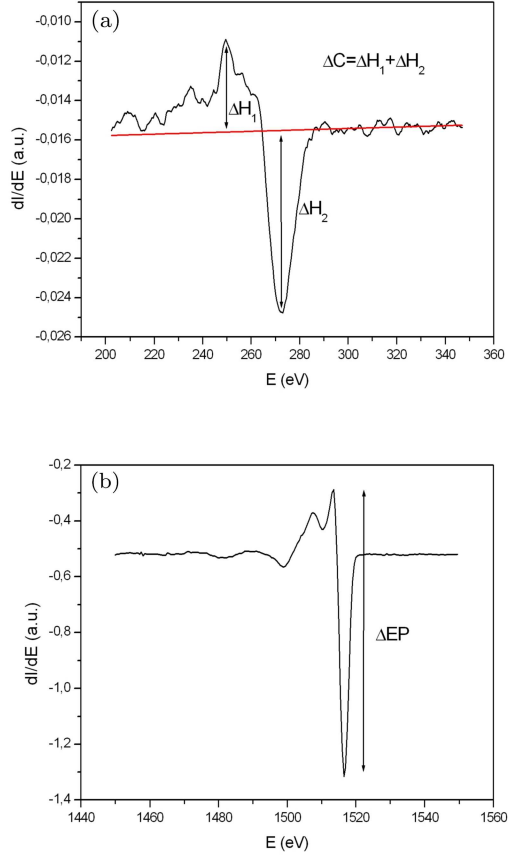}
 		\caption{%
	        Intensity measurement of  (a) C$_{KLL}$ Auger peak and  (b) the elastic peak. $I_C\propto\Delta C$ and $I_{EP}\propto\Delta EP$.
	     }%
	   \label{fig4}
 \end{figure}

The calibration was achieved for silicon samples coated with carbon layers up to 2.8 nm thick. The Auger spectra were recorded in the derivative mode, the Auger and elastic peak currents were evaluated from the peak-to-peak height deduced from derived spectra (see Fig. \ref{fig4}). This method is appropriate to relatively Gaussian shaped peaks (as shown in Ref. \onlinecite{Taylor1969}), which is the case in our study. For the carbon peak we have taken into account the base-line, as shown in figure \ref{fig4}.

During the calibration, the lock-in parameters of the Auger spectrometer were fixed: modulation amplitude and frequency were respectively 5 Vpp and 4.75 kHz. The time constant was set to 1 s, the integration time was 2 s and the lock-in used the High Dynamic Reserve mode.

\section{Results and discussion}
\begin{figure}
 \includegraphics[scale=1]{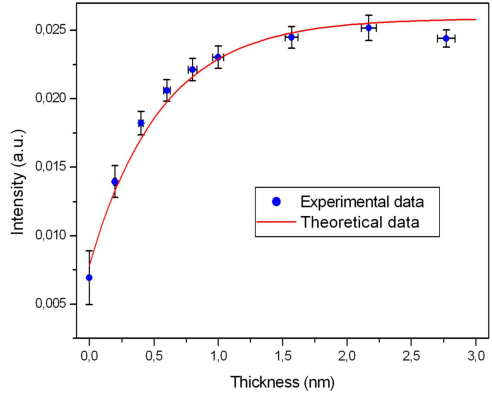}
 \caption{C$_{KLL}$ Auger peak intensity as a function of the carbon layer thickness.\label{fig5}}
 \end{figure}

\begin{figure}
 \includegraphics[scale=1]{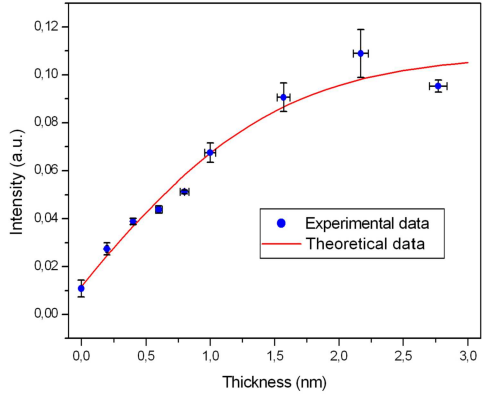}
 \caption{Ratio of the intensity of C$_{KLL}$ Auger peak and the elastic peak versus the carbon thickness.\label{fig6}}
 \end{figure}

At this point we observed the same behavior than that described in the Seah's model. However, we did not succeed in getting a clean surface free of carbon and the initial carbon contamination was modeled by a carbon layer of 0.2 nm.
By using the imfp of $C_{KLL}$ Auger electrons in carbon, $\lambda_{C,E_A}$ = 0.75 nm, we obtained (Fig \ref{fig5}) a correction factor of $k$ = 0.74, which is the value found by Seah\cite{Seah1972} for such a detector.
The relative intensity between the $C_{KLL}$ Auger peak and the elastic peak shows the trend described by Eq. \ref{equation3}. The theoretical curve shown in figure \ref{fig6} is obtained by using the following parameters: $\lambda_{C,E_A}$ = 0.75 nm, $\lambda_{C,EP}$ = 2.09 nm (for 1.5 keV electrons, according to the universal function given by Seah and Dench\cite{Mroz1994}), $k$ = 0.74, $R_{C}$=1.7 10$^{-4}$ and $R_{S}$=5 10$^{-4}$ from Ref.\onlinecite{Eckertova1990}.  The fit of Eq. \ref{equation3} to the experimental data provides $\alpha$=2$\times$10$^{-5}$. The measurable can be evaluated to a layer thickness spanning 0 to 1.5 nm.

\begin{figure}
 \includegraphics[scale=1]{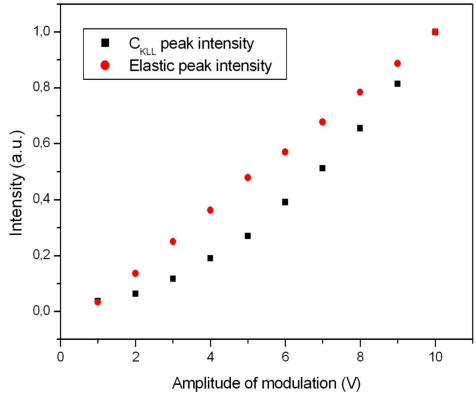}
 \caption{Normalized intensity of C$_{KLL}$ peak and the elastic peak as a function of the amplitude of modulation of the lock-in detection.\label{fig7}}
 \end{figure}

Different sources of uncertainties can affect this calibration. The two main sources come from the uncertainty on the thickness of the layer and on the intensity of the Auger and elastic peaks.
The uncertainty on the thickness of the layer is related to the measurement of the flux and its stability. The quartz crystal micro balance has a systematic error due to the uncertainty on the carbon density and on the geometric configuration. These uncertainties can be neglected. The stability of the flux was estimated to be 2\%.
The uncertainty on the peak amplitude measurement was obtained by statistics. As the number of Auger spectra is limited by the measurement time, this uncertainty was estimated from the standard deviation of the measurement corrected by a Student factor corresponding to a confidence interval of 70\%.
As all the measurements were performed by using specific lock-in parameters, the calibration can be considered valid for this set of parameters. However, some parameters such as the modulation amplitude, have to be tuned in order to improve the resolution. A correction factor was then applied to estimate the peak corresponding to a 5 Vpp modulation. Figure \ref{fig7} shows the evolution of the peak intensities as a function of the amplitude of modulation.

\section{Preliminary experiments}
\begin{figure}
 \includegraphics[scale=1]{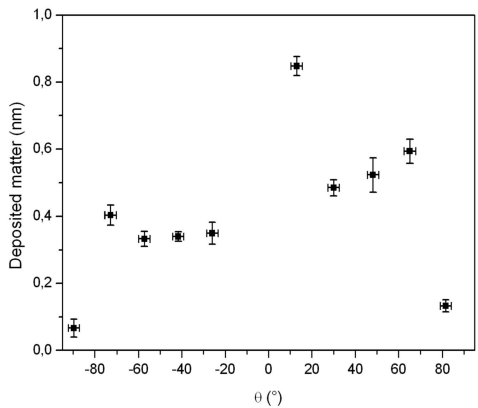}
 \caption{Distribution of the carbon deposition for a HOPG target irradiated with Ar$^{+}$ at 2.8 keV.\label{fig8}}
 \end{figure}

A set of experiments were performed by irradiating multiple fusion relevant targets. A first experiment was achieved with a target of highly oriented pyrolytic graphite (HOPG) irradiated by an Ar$^{+}$ beam of 1 $\mu$A  at 2.8 keV produced by a filament ion source. The target was previously cleaned under atmospheric pressure, and then sputtered under UHV by using the same ion beam than for the experiment. During this phase, the target was electrically biased in order to collect the secondary electrons emitted from the target. The current due to this emission is evaluated to be 10.8\%. During the actual sputtering experiment the target was grounded and the ion beam current was deduced analytically. Some extra Si catchers were placed away from the target in order to estimate the pollution during the irradiation. No pollution was detected from these catchers.
The catcher analysis started when a fluence of 1.91$\times$10$^{18}$ ions was reached. Figure \ref{fig8} shows the angular distribution of the deposition thickness, 0$^\circ$  corresponds to the beam incidence. The maximum is reached at around 13$^\circ$. It corresponds to a deposited carbon layer thickness of about 8.5 \AA. The distribution also shows two shoulders around 70$^\circ$ from the normal.
Tripathi et al\cite{Tripathi2003,Tripathi2008} showed similar results for HOPG irradiated at higher energy (Ag and Au at hundred MeV). They concluded by referring to a crystalline effect leading to preferential angles of sputtering. More data are however necessary to extrapolate such an explanation to lower energies.

\begin{figure}
 \includegraphics[scale=1]{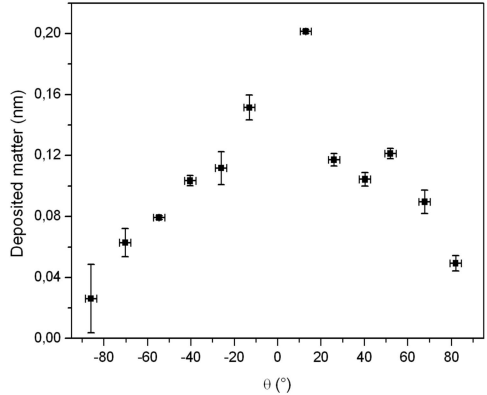}
 \caption{Distribution of the tungsten deposition for a tungsten target irradiated with Ar$^{+}$ at 7 keV.\label{fig9}}
 \end{figure}

A second experiment was performed with a target of tungsten irradiated by an Ar$^{+}$ beam of 1.5 $\mu$A  at 7 keV produced by the Mono1000 ion source. The fluence was 4$\times$10$^{16}$ ions. The layer thickness was measured by using Eqs \ref{equation1} and \ref{equation2} with the amplitude of the $W_{NVV}$ Auger electron (180 eV) and the elastic peak. Figure \ref{fig9} shows the distribution of the collected tungsten. The shape of the angular distribution of the sputtering yield exhibits a maximum near the normal incidence and can be described by a cosine law with extra preferential angle around 0$^\circ$ and 50$^\circ$. This observation was also previously made by Emmoth\cite{Emmoth1981} using similar conditions (tungsten bombarded by Ar$^+$ at 40 keV). 

\section{Conclusion and Perspectives}
An experimental setup that allows the measurement of the angular sputtering yield has been developed. It uses the so-called catcher method to collect sputtered particles, which are then analyzed by Auger electron spectroscopy without leaving the vacuum. This \textit{in situ} measurement allows the study of carbon based materials and other materials like tungsten.The Auger spectrometer was calibrated and preliminary experiments were preformed with HOPG and W targets irradiated by an Ar$^{+}$ beam, showing an angular yield distribution with preferential angles.

In the future, several experiments are scheduled in order to simulate more closely the irradiation conditions encountered in fusion reactors. For instance, we plan to irradiate various relevant plasma facing materials (such as CFC and W tiles) with very light ions (such as H$^+$, D$^+$ and He$^+$) at energies of a few hundreds eV. With this setup, it is also possible to study the effect of nitrogen seeding in fusion reactors\cite{Rapp2005}.  

\begin{acknowledgments}
We thank the F\'ed\'eration nationale de Recherche Fusion par Confinement Magn\'etique for their financial support.
\end{acknowledgments}


\begin{thebibliography}{24}%
\makeatletter
\providecommand \@ifxundefined [1]{%
 \@ifx{#1\undefined}
}%
\providecommand \@ifnum [1]{%
 \ifnum #1\expandafter \@firstoftwo
 \else \expandafter \@secondoftwo
 \fi
}%
\providecommand \@ifx [1]{%
 \ifx #1\expandafter \@firstoftwo
 \else \expandafter \@secondoftwo
 \fi
}%
\providecommand \natexlab [1]{#1}%
\providecommand \enquote  [1]{``#1''}%
\providecommand \bibnamefont  [1]{#1}%
\providecommand \bibfnamefont [1]{#1}%
\providecommand \citenamefont [1]{#1}%
\providecommand \href@noop [0]{\@secondoftwo}%
\providecommand \href [0]{\begingroup \@sanitize@url \@href}%
\providecommand \@href[1]{\@@startlink{#1}\@@href}%
\providecommand \@@href[1]{\endgroup#1\@@endlink}%
\providecommand \@sanitize@url [0]{\catcode `\\12\catcode `\$12\catcode
  `\&12\catcode `\#12\catcode `\^12\catcode `\_12\catcode `\%12\relax}%
\providecommand \@@startlink[1]{}%
\providecommand \@@endlink[0]{}%
\providecommand \url  [0]{\begingroup\@sanitize@url \@url }%
\providecommand \@url [1]{\endgroup\@href {#1}{\urlprefix }}%
\providecommand \urlprefix  [0]{URL }%
\providecommand \Eprint [0]{\href }%
\providecommand \doibase [0]{http://dx.doi.org/}%
\providecommand \selectlanguage [0]{\@gobble}%
\providecommand \bibinfo  [0]{\@secondoftwo}%
\providecommand \bibfield  [0]{\@secondoftwo}%
\providecommand \translation [1]{[#1]}%
\providecommand \BibitemOpen [0]{}%
\providecommand \bibitemStop [0]{}%
\providecommand \bibitemNoStop [0]{.\EOS\space}%
\providecommand \EOS [0]{\spacefactor3000\relax}%
\providecommand \BibitemShut  [1]{\csname bibitem#1\endcsname}%
\let\auto@bib@innerbib\@empty

\bibitem [{\citenamefont {Smentkowski}(2000)}]{Smentkowski20001}%
  \BibitemOpen
  \bibfield  {author} {\bibinfo {author} {\bibfnamefont {V.~S.}\ \bibnamefont
  {Smentkowski}},\ }\href {\doibase 10.1016/S0079-6816(99)00021-0} {\bibfield
  {journal} {\bibinfo  {journal} {Prog. Surf. Sci.}\ }\textbf {\bibinfo
  {volume} {64}},\ \bibinfo {pages} {1 } (\bibinfo {year} {2000})}\BibitemShut
  {NoStop}%
\bibitem [{\citenamefont {Roth}\ \emph {et~al.}(2009)\citenamefont {Roth},
  \citenamefont {Tsitrone}, \citenamefont {Loarte}, \citenamefont {Loarer},
  \citenamefont {Counsell}, \citenamefont {Neu}, \citenamefont {Philipps},
  \citenamefont {Brezinsek}, \citenamefont {Lehnen}, \citenamefont {Coad},
  \citenamefont {Grisolia}, \citenamefont {Schmid}, \citenamefont {Krieger},
  \citenamefont {Kallenbach}, \citenamefont {Lipschultz}, \citenamefont
  {Doerner}, \citenamefont {Causey}, \citenamefont {Alimov}, \citenamefont
  {Shu}, \citenamefont {Ogorodnikova}, \citenamefont {Kirschner}, \citenamefont
  {Federici},\ and\ \citenamefont {Kukushkin}}]{Roth20091}%
  \BibitemOpen
  \bibfield  {author} {\bibinfo {author} {\bibfnamefont {J.}~\bibnamefont
  {Roth}}, \bibinfo {author} {\bibfnamefont {E.}~\bibnamefont {Tsitrone}},
  \bibinfo {author} {\bibfnamefont {A.}~\bibnamefont {Loarte}}, \bibinfo
  {author} {\bibfnamefont {T.}~\bibnamefont {Loarer}}, \bibinfo {author}
  {\bibfnamefont {G.}~\bibnamefont {Counsell}}, \bibinfo {author}
  {\bibfnamefont {R.}~\bibnamefont {Neu}}, \bibinfo {author} {\bibfnamefont
  {V.}~\bibnamefont {Philipps}}, \bibinfo {author} {\bibfnamefont
  {S.}~\bibnamefont {Brezinsek}}, \bibinfo {author} {\bibfnamefont
  {M.}~\bibnamefont {Lehnen}}, \bibinfo {author} {\bibfnamefont
  {P.}~\bibnamefont {Coad}}, \bibinfo {author} {\bibfnamefont {C.}~\bibnamefont
  {Grisolia}}, \bibinfo {author} {\bibfnamefont {K.}~\bibnamefont {Schmid}},
  \bibinfo {author} {\bibfnamefont {K.}~\bibnamefont {Krieger}}, \bibinfo
  {author} {\bibfnamefont {A.}~\bibnamefont {Kallenbach}}, \bibinfo {author}
  {\bibfnamefont {B.}~\bibnamefont {Lipschultz}}, \bibinfo {author}
  {\bibfnamefont {R.}~\bibnamefont {Doerner}}, \bibinfo {author} {\bibfnamefont
  {R.}~\bibnamefont {Causey}}, \bibinfo {author} {\bibfnamefont
  {V.}~\bibnamefont {Alimov}}, \bibinfo {author} {\bibfnamefont
  {W.}~\bibnamefont {Shu}}, \bibinfo {author} {\bibfnamefont {O.}~\bibnamefont
  {Ogorodnikova}}, \bibinfo {author} {\bibfnamefont {A.}~\bibnamefont
  {Kirschner}}, \bibinfo {author} {\bibfnamefont {G.}~\bibnamefont {Federici}},
  \ and\ \bibinfo {author} {\bibfnamefont {A.}~\bibnamefont {Kukushkin}},\
  }\href {\doibase 10.1016/j.jnucmat.2009.01.037} {\bibfield  {journal}
  {\bibinfo  {journal} {J. Nucl. Mater.}\ }\textbf {\bibinfo {volume}
  {390-391}},\ \bibinfo {pages} {1 } (\bibinfo {year} {2009})}\BibitemShut
  {NoStop}%
\bibitem [{\citenamefont {Pitts}\ \emph {et~al.}(2005)\citenamefont {Pitts},
  \citenamefont {Coad}, \citenamefont {Coster}, \citenamefont {Federici},
  \citenamefont {Fundamenski}, \citenamefont {Horacek}, \citenamefont
  {Krieger}, \citenamefont {Kukushkin}, \citenamefont {Likonen}, \citenamefont
  {Matthews}, \citenamefont {Rubel}, \citenamefont {Strachan},\ and\
  \citenamefont {Contributors}}]{Pitts2005}%
  \BibitemOpen
  \bibfield  {author} {\bibinfo {author} {\bibfnamefont {R.}~\bibnamefont
  {Pitts}}, \bibinfo {author} {\bibfnamefont {J.}~\bibnamefont {Coad}},
  \bibinfo {author} {\bibfnamefont {D.}~\bibnamefont {Coster}}, \bibinfo
  {author} {\bibfnamefont {G.}~\bibnamefont {Federici}}, \bibinfo {author}
  {\bibfnamefont {W.}~\bibnamefont {Fundamenski}}, \bibinfo {author}
  {\bibfnamefont {J.}~\bibnamefont {Horacek}}, \bibinfo {author} {\bibfnamefont
  {K.}~\bibnamefont {Krieger}}, \bibinfo {author} {\bibfnamefont
  {A.}~\bibnamefont {Kukushkin}}, \bibinfo {author} {\bibfnamefont
  {J.}~\bibnamefont {Likonen}}, \bibinfo {author} {\bibfnamefont
  {G.}~\bibnamefont {Matthews}}, \bibinfo {author} {\bibfnamefont
  {M.}~\bibnamefont {Rubel}}, \bibinfo {author} {\bibfnamefont
  {J.}~\bibnamefont {Strachan}}, \ and\ \bibinfo {author} {\bibfnamefont
  {J.-E.}\ \bibnamefont {Contributors}},\ }\href {\doibase
  {10.1088/0741-3335/47/12B/S22}} {\bibfield  {journal} {\bibinfo  {journal}
  {Plasma. Phys. Contr. F.}\ }\textbf {\bibinfo {volume} {47}},\ \bibinfo
  {pages} {303} (\bibinfo {year} {2005})}\BibitemShut {NoStop}%
\bibitem [{\citenamefont {Federici}\ \emph {et~al.}(2001)\citenamefont
  {Federici}, \citenamefont {Skinner}, \citenamefont {Brooks}, \citenamefont
  {Coad}, \citenamefont {Grisolia}, \citenamefont {Haasz}, \citenamefont
  {Hassanein}, \citenamefont {Philipps}, \citenamefont {Pitcher}, \citenamefont
  {Roth}, \citenamefont {Wampler},\ and\ \citenamefont {Whyte}}]{Federici2001}%
  \BibitemOpen
  \bibfield  {author} {\bibinfo {author} {\bibfnamefont {G.}~\bibnamefont
  {Federici}}, \bibinfo {author} {\bibfnamefont {C.}~\bibnamefont {Skinner}},
  \bibinfo {author} {\bibfnamefont {J.}~\bibnamefont {Brooks}}, \bibinfo
  {author} {\bibfnamefont {J.}~\bibnamefont {Coad}}, \bibinfo {author}
  {\bibfnamefont {C.}~\bibnamefont {Grisolia}}, \bibinfo {author}
  {\bibfnamefont {A.}~\bibnamefont {Haasz}}, \bibinfo {author} {\bibfnamefont
  {A.}~\bibnamefont {Hassanein}}, \bibinfo {author} {\bibfnamefont
  {V.}~\bibnamefont {Philipps}}, \bibinfo {author} {\bibfnamefont
  {C.}~\bibnamefont {Pitcher}}, \bibinfo {author} {\bibfnamefont
  {J.}~\bibnamefont {Roth}}, \bibinfo {author} {\bibfnamefont {W.}~\bibnamefont
  {Wampler}}, \ and\ \bibinfo {author} {\bibfnamefont {D.}~\bibnamefont
  {Whyte}},\ }\href {http://stacks.iop.org/0029-5515/41/i=12/a=218} {\bibfield
  {journal} {\bibinfo  {journal} {Nucl. Fusion}\ }\textbf {\bibinfo {volume}
  {41}},\ \bibinfo {pages} {1967} (\bibinfo {year} {2001})}\BibitemShut
  {NoStop}%
\bibitem [{\citenamefont {Goebel}\ \emph {et~al.}(1987)\citenamefont {Goebel},
  \citenamefont {Hirooka}, \citenamefont {Conn}, \citenamefont {Leung},
  \citenamefont {Campbell}, \citenamefont {Bohdansky}, \citenamefont {Wilson},
  \citenamefont {Bauer}, \citenamefont {Causey}, \citenamefont {Pontau},
  \citenamefont {Krauss}, \citenamefont {Gruen},\ and\ \citenamefont
  {Mendelsohn}}]{Goebel198761}%
  \BibitemOpen
  \bibfield  {author} {\bibinfo {author} {\bibfnamefont {D.}~\bibnamefont
  {Goebel}}, \bibinfo {author} {\bibfnamefont {Y.}~\bibnamefont {Hirooka}},
  \bibinfo {author} {\bibfnamefont {R.}~\bibnamefont {Conn}}, \bibinfo {author}
  {\bibfnamefont {W.}~\bibnamefont {Leung}}, \bibinfo {author} {\bibfnamefont
  {G.}~\bibnamefont {Campbell}}, \bibinfo {author} {\bibfnamefont
  {J.}~\bibnamefont {Bohdansky}}, \bibinfo {author} {\bibfnamefont
  {K.}~\bibnamefont {Wilson}}, \bibinfo {author} {\bibfnamefont
  {W.}~\bibnamefont {Bauer}}, \bibinfo {author} {\bibfnamefont
  {R.}~\bibnamefont {Causey}}, \bibinfo {author} {\bibfnamefont
  {A.}~\bibnamefont {Pontau}}, \bibinfo {author} {\bibfnamefont
  {A.}~\bibnamefont {Krauss}}, \bibinfo {author} {\bibfnamefont
  {D.}~\bibnamefont {Gruen}}, \ and\ \bibinfo {author} {\bibfnamefont
  {M.}~\bibnamefont {Mendelsohn}},\ }\href {\doibase
  10.1016/0022-3115(87)90310-2} {\bibfield  {journal} {\bibinfo  {journal} {J.
  Nucl. Mater.}\ }\textbf {\bibinfo {volume} {145-147}},\ \bibinfo {pages} {61
  } (\bibinfo {year} {1987})}\BibitemShut {NoStop}%
\bibitem [{\citenamefont {Aumayr}\ and\ \citenamefont
  {Winter}(2004)}]{Aumayr2004}%
  \BibitemOpen
  \bibfield  {author} {\bibinfo {author} {\bibfnamefont {F.}~\bibnamefont
  {Aumayr}}\ and\ \bibinfo {author} {\bibfnamefont {H.}~\bibnamefont
  {Winter}},\ }\href {\doibase {10.1098/rsta.2003.1300}} {\bibfield  {journal}
  {\bibinfo  {journal} {Philos. T. Roy. Soc. A.}\ }\textbf {\bibinfo {volume}
  {{362}}},\ \bibinfo {pages} {77} (\bibinfo {year} {{2004}})}\BibitemShut
  {NoStop}%
\bibitem [{\citenamefont {Keim}\ \emph {et~al.}(2011)\citenamefont {Keim},
  \citenamefont {Rasul}, \citenamefont {Endstrasser}, \citenamefont {Scheier},
  \citenamefont {Mark},\ and\ \citenamefont {Herman}}]{Keim2011}%
  \BibitemOpen
  \bibfield  {author} {\bibinfo {author} {\bibfnamefont {A.}~\bibnamefont
  {Keim}}, \bibinfo {author} {\bibfnamefont {B.}~\bibnamefont {Rasul}},
  \bibinfo {author} {\bibfnamefont {N.}~\bibnamefont {Endstrasser}}, \bibinfo
  {author} {\bibfnamefont {P.}~\bibnamefont {Scheier}}, \bibinfo {author}
  {\bibfnamefont {T.~D.}\ \bibnamefont {Mark}}, \ and\ \bibinfo {author}
  {\bibfnamefont {Z.}~\bibnamefont {Herman}},\ }\href {<Go to
  ISI>://000295351200016} {\bibfield  {journal} {\bibinfo  {journal} {Int. J.
  Mass Spectrom.}\ }\textbf {\bibinfo {volume} {306}},\ \bibinfo {pages} {204}
  (\bibinfo {year} {2011})}\BibitemShut {NoStop}%
\bibitem [{\citenamefont {Bouffard}\ \emph {et~al.}(1998)\citenamefont
  {Bouffard}, \citenamefont {Duraud}, \citenamefont {Mosbah},\ and\
  \citenamefont {Schlutig}}]{Bouffard1998372}%
  \BibitemOpen
  \bibfield  {author} {\bibinfo {author} {\bibfnamefont {S.}~\bibnamefont
  {Bouffard}}, \bibinfo {author} {\bibfnamefont {J.}~\bibnamefont {Duraud}},
  \bibinfo {author} {\bibfnamefont {M.}~\bibnamefont {Mosbah}}, \ and\ \bibinfo
  {author} {\bibfnamefont {S.}~\bibnamefont {Schlutig}},\ }\href {\doibase
  10.1016/S0168-583X(98)00170-0} {\bibfield  {journal} {\bibinfo  {journal}
  {Nucl. Instrum. Meth. B.}\ }\textbf {\bibinfo {volume} {141}},\ \bibinfo
  {pages} {372} (\bibinfo {year} {1998})}\BibitemShut {NoStop}%
\bibitem [{\citenamefont {Ban-d'Etat}\ \emph {et~al.}(2004)\citenamefont
  {Ban-d'Etat}, \citenamefont {Haranger}, \citenamefont {Boduch}, \citenamefont
  {Bouffard}, \citenamefont {Lebius}, \citenamefont {Maunoury}, \citenamefont
  {Pacquet}, \citenamefont {Rothard}, \citenamefont {Clerc}, \citenamefont
  {Garrido}, \citenamefont {Thom\'{e}}, \citenamefont {Hellhammer},
  \citenamefont {Pe{\v s}i\'{c}},\ and\ \citenamefont
  {Stolterfoht}}]{Etat2004}%
  \BibitemOpen
  \bibfield  {author} {\bibinfo {author} {\bibfnamefont {B.}~\bibnamefont
  {Ban-d'Etat}}, \bibinfo {author} {\bibfnamefont {F.}~\bibnamefont
  {Haranger}}, \bibinfo {author} {\bibfnamefont {P.}~\bibnamefont {Boduch}},
  \bibinfo {author} {\bibfnamefont {S.}~\bibnamefont {Bouffard}}, \bibinfo
  {author} {\bibfnamefont {H.}~\bibnamefont {Lebius}}, \bibinfo {author}
  {\bibfnamefont {L.}~\bibnamefont {Maunoury}}, \bibinfo {author}
  {\bibfnamefont {J.~Y.}\ \bibnamefont {Pacquet}}, \bibinfo {author}
  {\bibfnamefont {H.}~\bibnamefont {Rothard}}, \bibinfo {author} {\bibfnamefont
  {C.}~\bibnamefont {Clerc}}, \bibinfo {author} {\bibfnamefont
  {F.}~\bibnamefont {Garrido}}, \bibinfo {author} {\bibfnamefont
  {L.}~\bibnamefont {Thom\'{e}}}, \bibinfo {author} {\bibfnamefont
  {R.}~\bibnamefont {Hellhammer}}, \bibinfo {author} {\bibfnamefont
  {Z.}~\bibnamefont {Pe{\v s}i\'{c}}}, \ and\ \bibinfo {author} {\bibfnamefont
  {N.}~\bibnamefont {Stolterfoht}},\ }\href
  {http://stacks.iop.org/1402-4896/2004/i=T110/a=069} {\bibfield  {journal}
  {\bibinfo  {journal} {Phys. Scr.}\ }\textbf {\bibinfo {volume} {2004}},\
  \bibinfo {pages} {389} (\bibinfo {year} {2004})}\BibitemShut {NoStop}%
\bibitem [{\citenamefont {Tripathi}\ \emph {et~al.}(2003)\citenamefont
  {Tripathi}, \citenamefont {Khan}, \citenamefont {Srivastava}, \citenamefont
  {Kumar}, \citenamefont {Kumar}, \citenamefont {Rao}, \citenamefont {Lakshmi},
  \citenamefont {Siddiqui}, \citenamefont {Bajwa}, \citenamefont {Nagaraja},
  \citenamefont {Mittal}, \citenamefont {Szokefalvi}, \citenamefont {Kurth},
  \citenamefont {Pandey}, \citenamefont {Avasthi},\ and\ \citenamefont
  {Carstanjen}}]{Tripathi2003}%
  \BibitemOpen
  \bibfield  {author} {\bibinfo {author} {\bibfnamefont {A.}~\bibnamefont
  {Tripathi}}, \bibinfo {author} {\bibfnamefont {S.~A.}\ \bibnamefont {Khan}},
  \bibinfo {author} {\bibfnamefont {S.~K.}\ \bibnamefont {Srivastava}},
  \bibinfo {author} {\bibfnamefont {M.}~\bibnamefont {Kumar}}, \bibinfo
  {author} {\bibfnamefont {S.}~\bibnamefont {Kumar}}, \bibinfo {author}
  {\bibfnamefont {S.~V. S.~N.}\ \bibnamefont {Rao}}, \bibinfo {author}
  {\bibfnamefont {G.~B. V.~S.}\ \bibnamefont {Lakshmi}}, \bibinfo {author}
  {\bibfnamefont {A.~M.}\ \bibnamefont {Siddiqui}}, \bibinfo {author}
  {\bibfnamefont {N.}~\bibnamefont {Bajwa}}, \bibinfo {author} {\bibfnamefont
  {H.~S.}\ \bibnamefont {Nagaraja}}, \bibinfo {author} {\bibfnamefont {V.~K.}\
  \bibnamefont {Mittal}}, \bibinfo {author} {\bibfnamefont {A.}~\bibnamefont
  {Szokefalvi}}, \bibinfo {author} {\bibfnamefont {M.}~\bibnamefont {Kurth}},
  \bibinfo {author} {\bibfnamefont {A.~C.}\ \bibnamefont {Pandey}}, \bibinfo
  {author} {\bibfnamefont {D.~K.}\ \bibnamefont {Avasthi}}, \ and\ \bibinfo
  {author} {\bibfnamefont {H.~D.}\ \bibnamefont {Carstanjen}},\ }\href
  {http://www.sciencedirect.com/science/article/pii/S0168583X03017427}
  {\bibfield  {journal} {\bibinfo  {journal} {Nucl. Instrum. Meth. B.}\
  }\textbf {\bibinfo {volume} {212}},\ \bibinfo {pages} {402} (\bibinfo {year}
  {2003})}\BibitemShut {NoStop}%
\bibitem [{\citenamefont {Tripathi}\ \emph {et~al.}(2008)\citenamefont
  {Tripathi}, \citenamefont {Khan}, \citenamefont {Kumar}, \citenamefont
  {Baranwal}, \citenamefont {Krishna}, \citenamefont {Kumar}, \citenamefont
  {Pandey},\ and\ \citenamefont {Avasthi}}]{Tripathi2008}%
  \BibitemOpen
  \bibfield  {author} {\bibinfo {author} {\bibfnamefont {A.}~\bibnamefont
  {Tripathi}}, \bibinfo {author} {\bibfnamefont {S.~A.}\ \bibnamefont {Khan}},
  \bibinfo {author} {\bibfnamefont {M.}~\bibnamefont {Kumar}}, \bibinfo
  {author} {\bibfnamefont {V.}~\bibnamefont {Baranwal}}, \bibinfo {author}
  {\bibfnamefont {R.}~\bibnamefont {Krishna}}, \bibinfo {author} {\bibfnamefont
  {S.}~\bibnamefont {Kumar}}, \bibinfo {author} {\bibfnamefont {A.~C.}\
  \bibnamefont {Pandey}}, \ and\ \bibinfo {author} {\bibfnamefont {D.~K.}\
  \bibnamefont {Avasthi}},\ }\href
  {http://www.sciencedirect.com/science/article/pii/S0168583X08000591}
  {\bibfield  {journal} {\bibinfo  {journal} {Nucl. Instrum. Meth. B.}\
  }\textbf {\bibinfo {volume} {266}},\ \bibinfo {pages} {1265} (\bibinfo {year}
  {2008})}\BibitemShut {NoStop}%
\bibitem [{\citenamefont {Roth}\ and\ \citenamefont {Hopf}(2004)}]{Roth2004}%
  \BibitemOpen
  \bibfield  {author} {\bibinfo {author} {\bibfnamefont {J.}~\bibnamefont
  {Roth}}\ and\ \bibinfo {author} {\bibfnamefont {C.}~\bibnamefont {Hopf}},\
  }\href {http://www.sciencedirect.com/science/article/pii/S0022311504004878}
  {\bibfield  {journal} {\bibinfo  {journal} {J. Nucl. Mater.}\ }\textbf
  {\bibinfo {volume} {334}},\ \bibinfo {pages} {97} (\bibinfo {year}
  {2004})}\BibitemShut {NoStop}%
\bibitem [{\citenamefont {Pellerin}\ and\ \citenamefont
  {Le~Gressus}(1979)}]{Pellerin1979}%
  \BibitemOpen
  \bibfield  {author} {\bibinfo {author} {\bibfnamefont {F.}~\bibnamefont
  {Pellerin}}\ and\ \bibinfo {author} {\bibfnamefont {C.}~\bibnamefont
  {Le~Gressus}},\ }\href
  {http://www.sciencedirect.com/science/article/pii/0039602879901791}
  {\bibfield  {journal} {\bibinfo  {journal} {Surf. Sci.}\ }\textbf {\bibinfo
  {volume} {87}},\ \bibinfo {pages} {203} (\bibinfo {year} {1979})}\BibitemShut
  {NoStop}%
\bibitem [{\citenamefont {Pierret}\ \emph {et~al.}(2008)\citenamefont
  {Pierret}, \citenamefont {Maunoury}, \citenamefont {Biri}, \citenamefont
  {Pacquet}, \citenamefont {Tuske},\ and\ \citenamefont
  {Delferriere}}]{Pierret2008}%
  \BibitemOpen
  \bibfield  {author} {\bibinfo {author} {\bibfnamefont {C.}~\bibnamefont
  {Pierret}}, \bibinfo {author} {\bibfnamefont {L.}~\bibnamefont {Maunoury}},
  \bibinfo {author} {\bibfnamefont {S.}~\bibnamefont {Biri}}, \bibinfo {author}
  {\bibfnamefont {J.~Y.}\ \bibnamefont {Pacquet}}, \bibinfo {author}
  {\bibfnamefont {O.}~\bibnamefont {Tuske}}, \ and\ \bibinfo {author}
  {\bibfnamefont {O.}~\bibnamefont {Delferriere}},\ }\href
  {http://dx.doi.org/10.1063/1.2814260} {\bibfield  {journal} {\bibinfo
  {journal} {Rev. Sci. Instrum.}\ }\textbf {\bibinfo {volume} {79}},\ \bibinfo
  {pages} {02B703} (\bibinfo {year} {2008})}\BibitemShut {NoStop}%
\bibitem [{\citenamefont {Lebius}\ \emph {et~al.}(2003)\citenamefont {Lebius},
  \citenamefont {Brenac}, \citenamefont {Huber}, \citenamefont {Maunoury},
  \citenamefont {Gustavo},\ and\ \citenamefont {Cormier}}]{Lebius2003}%
  \BibitemOpen
  \bibfield  {author} {\bibinfo {author} {\bibfnamefont {H.}~\bibnamefont
  {Lebius}}, \bibinfo {author} {\bibfnamefont {A.}~\bibnamefont {Brenac}},
  \bibinfo {author} {\bibfnamefont {B.~A.}\ \bibnamefont {Huber}}, \bibinfo
  {author} {\bibfnamefont {L.}~\bibnamefont {Maunoury}}, \bibinfo {author}
  {\bibfnamefont {F.}~\bibnamefont {Gustavo}}, \ and\ \bibinfo {author}
  {\bibfnamefont {D.}~\bibnamefont {Cormier}},\ }\href
  {http://dx.doi.org/10.1063/1.1556952} {\bibfield  {journal} {\bibinfo
  {journal} {Rev Sci Instrum}\ }\textbf {\bibinfo {volume} {74}},\ \bibinfo
  {pages} {2276} (\bibinfo {year} {2003})}\BibitemShut {NoStop}%
\bibitem [{\citenamefont {Castle}(1983)}]{Briggs1983}%
  \BibitemOpen
  \bibfield  {author} {\bibinfo {author} {\bibfnamefont {J.~E.}\ \bibnamefont
  {Castle}},\ }\href@noop {} {\emph {\bibinfo {title} {Practical surface
  analysis by Auger and X-ray photoelectron spectroscopy.}}},\ edited by\
  \bibinfo {editor} {\bibfnamefont {D.}~\bibnamefont {Briggs}}\ and\ \bibinfo
  {editor} {\bibfnamefont {M.~P.}\ \bibnamefont {Seah}}\ (\bibinfo  {publisher}
  {John Wiley and Sons Ltd, Chichester},\ \bibinfo {year} {1983})\BibitemShut
  {NoStop}%
\bibitem [{\citenamefont {Argile}\ and\ \citenamefont
  {Rhead}(1989)}]{Argile1989}%
  \BibitemOpen
  \bibfield  {author} {\bibinfo {author} {\bibfnamefont {C.}~\bibnamefont
  {Argile}}\ and\ \bibinfo {author} {\bibfnamefont {G.~E.}\ \bibnamefont
  {Rhead}},\ }\href
  {http://www.sciencedirect.com/science/article/pii/0167572989900010}
  {\bibfield  {journal} {\bibinfo  {journal} {Surf. Sci. Rep.}\ }\textbf
  {\bibinfo {volume} {10}},\ \bibinfo {pages} {277} (\bibinfo {year}
  {1989})}\BibitemShut {NoStop}%
\bibitem [{\citenamefont {Akc\"{o}ltekin}\ \emph {et~al.}(2009)\citenamefont
  {Akc\"{o}ltekin}, \citenamefont {Alzaher}, \citenamefont {Ban~d'Etat},
  \citenamefont {Been}, \citenamefont {Boduch}, \citenamefont {Cassimi},
  \citenamefont {Hijazi}, \citenamefont {Lebius}, \citenamefont {Manil},
  \citenamefont {Ramillon}, \citenamefont {Rothard}, \citenamefont
  {Schleberger},\ and\ \citenamefont {da~Silveira}}]{Akcoeltekin2009}%
  \BibitemOpen
  \bibfield  {author} {\bibinfo {author} {\bibfnamefont {S.}~\bibnamefont
  {Akc\"{o}ltekin}}, \bibinfo {author} {\bibfnamefont {I.}~\bibnamefont
  {Alzaher}}, \bibinfo {author} {\bibfnamefont {B.}~\bibnamefont {Ban~d'Etat}},
  \bibinfo {author} {\bibfnamefont {T.}~\bibnamefont {Been}}, \bibinfo {author}
  {\bibfnamefont {P.}~\bibnamefont {Boduch}}, \bibinfo {author} {\bibfnamefont
  {A.}~\bibnamefont {Cassimi}}, \bibinfo {author} {\bibfnamefont
  {H.}~\bibnamefont {Hijazi}}, \bibinfo {author} {\bibfnamefont
  {H.}~\bibnamefont {Lebius}}, \bibinfo {author} {\bibfnamefont
  {B.}~\bibnamefont {Manil}}, \bibinfo {author} {\bibfnamefont {J.~M.}\
  \bibnamefont {Ramillon}}, \bibinfo {author} {\bibfnamefont {H.}~\bibnamefont
  {Rothard}}, \bibinfo {author} {\bibfnamefont {M.}~\bibnamefont
  {Schleberger}}, \ and\ \bibinfo {author} {\bibfnamefont {E.~F.}\ \bibnamefont
  {da~Silveira}},\ }\href
  {http://www.sciencedirect.com/science/article/pii/S0168583X08012615}
  {\bibfield  {journal} {\bibinfo  {journal} {Nucl. Instrum. Meth. B.}\
  }\textbf {\bibinfo {volume} {267}},\ \bibinfo {pages} {649} (\bibinfo {year}
  {2009})}\BibitemShut {NoStop}%
\bibitem [{\citenamefont {Taylor}(1969)}]{Taylor1969}%
  \BibitemOpen
  \bibfield  {author} {\bibinfo {author} {\bibfnamefont {N.~J.}\ \bibnamefont
  {Taylor}},\ }\href {http://dx.doi.org/10.1063/1.1684071} {\bibfield
  {journal} {\bibinfo  {journal} {Rev. Sci. Instrum.}\ }\textbf {\bibinfo
  {volume} {40}},\ \bibinfo {pages} {792} (\bibinfo {year} {1969})}\BibitemShut
  {NoStop}%
\bibitem [{\citenamefont {Seah}(1972)}]{Seah1972}%
  \BibitemOpen
  \bibfield  {author} {\bibinfo {author} {\bibfnamefont {M.~P.}\ \bibnamefont
  {Seah}},\ }\href
  {http://www.sciencedirect.com/science/article/pii/0039602872901963}
  {\bibfield  {journal} {\bibinfo  {journal} {Surf. Sci.}\ }\textbf {\bibinfo
  {volume} {32}},\ \bibinfo {pages} {703} (\bibinfo {year} {1972})}\BibitemShut
  {NoStop}%
\bibitem [{\citenamefont {Mr\"{o}z}(1994)}]{Mroz1994}%
  \BibitemOpen
  \bibfield  {author} {\bibinfo {author} {\bibfnamefont {S.}~\bibnamefont
  {Mr\"{o}z}},\ }\href
  {http://www.sciencedirect.com/science/article/pii/0079681694900167}
  {\bibfield  {journal} {\bibinfo  {journal} {Prog. Surf. Sci.}\ }\textbf
  {\bibinfo {volume} {46}},\ \bibinfo {pages} {377} (\bibinfo {year}
  {1994})}\BibitemShut {NoStop}%
\bibitem [{\citenamefont {Eckertova}\ and\ \citenamefont
  {Kleint}(1990)}]{Eckertova1990}%
  \BibitemOpen
  \bibfield  {author} {\bibinfo {author} {\bibfnamefont {L.}~\bibnamefont
  {Eckertova}}\ and\ \bibinfo {author} {\bibfnamefont {C.}~\bibnamefont
  {Kleint}},\ }\href
  {http://www.sciencedirect.com/science/article/pii/003960289090710P}
  {\bibfield  {journal} {\bibinfo  {journal} {Surf. Sci.}\ }\textbf {\bibinfo
  {volume} {231}},\ \bibinfo {pages} {168} (\bibinfo {year}
  {1990})}\BibitemShut {NoStop}%
\bibitem [{\citenamefont {Emmoth}(1981)}]{Emmoth1981}%
  \BibitemOpen
  \bibfield  {author} {\bibinfo {author} {\bibfnamefont {B.}~\bibnamefont
  {Emmoth}},\ }\href {http://stacks.iop.org/1402-4896/24/i=3/a=017} {\bibfield
  {journal} {\bibinfo  {journal} {Phys. Scr.}\ }\textbf {\bibinfo {volume}
  {24}},\ \bibinfo {pages} {609} (\bibinfo {year} {1981})}\BibitemShut
  {NoStop}%
\bibitem [{\citenamefont {Rapp}\ \emph {et~al.}(2005)\citenamefont {Rapp},
  \citenamefont {Matthews}, \citenamefont {Monier-Garbet}, \citenamefont
  {Sartori}, \citenamefont {Corre}, \citenamefont {Eich}, \citenamefont
  {Felton}, \citenamefont {Fundamenski}, \citenamefont {Giroud}, \citenamefont
  {Huber}, \citenamefont {Jachmich}, \citenamefont {Morgan}, \citenamefont
  {O'Mullane}, \citenamefont {Koslowski}, \citenamefont {Stamp},\ and\
  \citenamefont {contributors to the Efda-J. E. T.~work programme}}]{Rapp2005}%
  \BibitemOpen
  \bibfield  {author} {\bibinfo {author} {\bibfnamefont {J.}~\bibnamefont
  {Rapp}}, \bibinfo {author} {\bibfnamefont {G.~F.}\ \bibnamefont {Matthews}},
  \bibinfo {author} {\bibfnamefont {P.}~\bibnamefont {Monier-Garbet}}, \bibinfo
  {author} {\bibfnamefont {R.}~\bibnamefont {Sartori}}, \bibinfo {author}
  {\bibfnamefont {Y.}~\bibnamefont {Corre}}, \bibinfo {author} {\bibfnamefont
  {T.}~\bibnamefont {Eich}}, \bibinfo {author} {\bibfnamefont {R.}~\bibnamefont
  {Felton}}, \bibinfo {author} {\bibfnamefont {W.}~\bibnamefont {Fundamenski}},
  \bibinfo {author} {\bibfnamefont {C.}~\bibnamefont {Giroud}}, \bibinfo
  {author} {\bibfnamefont {A.}~\bibnamefont {Huber}}, \bibinfo {author}
  {\bibfnamefont {S.}~\bibnamefont {Jachmich}}, \bibinfo {author}
  {\bibfnamefont {P.}~\bibnamefont {Morgan}}, \bibinfo {author} {\bibfnamefont
  {M.}~\bibnamefont {O'Mullane}}, \bibinfo {author} {\bibfnamefont {H.~R.}\
  \bibnamefont {Koslowski}}, \bibinfo {author} {\bibfnamefont {M.}~\bibnamefont
  {Stamp}}, \ and\ \bibinfo {author} {\bibnamefont {contributors to the Efda-J.
  E. T.~work programme}},\ }\href
  {http://www.sciencedirect.com/science/article/pii/S0022311504009742}
  {\bibfield  {journal} {\bibinfo  {journal} {J. Nucl. Mater.}\ }\textbf
  {\bibinfo {volume} {337-339}},\ \bibinfo {pages} {826} (\bibinfo {year}
  {2005})}\BibitemShut {NoStop}%

\end{thebibliography}

%

\end{document}